\documentclass[reprint,amsmath,amssymb,aps, longbibliography]{revtex4-1}
\usepackage{graphicx}
\usepackage[centerlast]{caption}
\usepackage{subcaption}

\usepackage{natbib}
\usepackage{multirow}

\usepackage{array}
\newcolumntype{L}[1]{>{\raggedright\let\newline\\\arraybackslash\hspace{0pt}}m{#1}}
\newcolumntype{C}[1]{>{\centering\let\newline\\\arraybackslash\hspace{0pt}}m{#1}}
\newcolumntype{R}[1]{>{\raggedleft\let\newline\\\arraybackslash\hspace{0pt}}m{#1}}

\usepackage{hyperref}
\usepackage{psfrag}
\usepackage{amsmath}
\usepackage{amssymb}
\usepackage{gensymb}

\makeatletter
\renewcommand\@make@capt@title[2]{%
 \@ifx@empty\float@link{\@firstofone}{\expandafter\href\expandafter{\float@link}}%
  {\textbf{#1}}\@caption@fignum@sep#2\quad
}%
\makeatother

\begin{document}

\title{Toolboxes and handing students a hammer: The effects of cueing and instruction on getting students to think critically}

\author{N.G. Holmes}
\email[]{ngholmes@cornell.edu}
\affiliation{Laboratory for Atomic and Solid State Physics, Department of Physics, Cornell University, Ithaca, NY}
\author{Dhaneesh Kumar}
\author{D.A. Bonn}

\affiliation{Department of Physics and Astronomy, University of British Columbia, Vancouver, BC}

\date{\today}

\begin{abstract}

Developing critical thinking skills is a common goal of an undergraduate physics curriculum. How do students make sense of evidence and what do they do with it? In this study, we evaluated students' critical thinking behaviors through their written notebooks in an introductory physics laboratory course. We compared student behaviors in the Structured Quantitative Inquiry Labs (SQILabs) curriculum to a control group and evaluated the fragility of these behaviors through procedural cueing. We found that the SQILabs were generally effective at improving the quality of students' reasoning about data and making decisions from data. These improvements in reasoning and sensemaking were thwarted, however, by a procedural cue. We describe these changes in behavior through the lens of epistemological frames and task orientation, invoked by the instructional moves. 

\end{abstract}

\maketitle

\section{Introduction}

A new curricular approach for introductory labs, the Structured Quantitative Inquiry Labs (SQILabs), aims to develop students' quantitative critical thinking skills. We define critical thinking as the process through which one decides what to believe. In this context we focus on `believing' data, evidence, and models. Previous evaluation found that the SQILabs structure dramatically improves the fraction of students who reflect on their results from an experiment, iterate to improve their measurements, and evaluate disagreements between models and data \cite{HolmesPNAS, HolmesTPT}. The particular mechanisms for these improvements, however, have not been thoroughly evaluated and it is unclear which elements of the SQILabs were particularly salient for causing the improvements. 

In this study, we aimed to explore the effects of various elements of the SQILabs on students' task orientation and epistemologies. We tested the effects of prior instruction (SQILabs versus a control group) and the effects of cueing with a new analysis tool. We infer students' task orientation and epistemologies from their experimentation behaviors in an unstructured lab activity.

In the following section, we expand upon the motivation for these research questions in the context of prior literature.

\section{Literature review}

Introductory physics lab courses are often service courses, intended to serve a diverse group of students. These courses often aim to provide students with a toolbox of skills and knowledge that will be useful to them in the future. The contents of this toolbox may differ widely between institutions \cite{Hofstein1982, Hofstein2004, ALR}. Whether the toolbox is one of skills to solve problems, physics concepts to help students reason about the physical world, or critical thinking skills to help students reason about data and models, the items in the toolbox alone are insufficient for preparing students for the future. That is, it is only useful to have a hammer if you know how and when to use it. 

In this work, we build from the notion that students' decisions to use particular tools will be related to their task orientation and their epistemological frame. We define epistemological frame as ``the student's perception or judgment of the kind of knowledge that is appropriate to bring to bear in a particular situation" \cite[][p.1]{Bing2009}. In this way, an epistemological frame is related to a students' task orientation, defined here as ``interpreting task demands and then direct[ing] further learning activities accordingly" \cite{Butler2004}. Students may approach a classroom task with a number of different frames. 

The reason(s) a student engages through a particular frame can depend on a number of variables including their prior instruction, the task instructions, their current social/emotional state, or their competing motivations. A students' frame can be seen as a function of the individual student (each bringing in a unique set of resources from their current and past experiences) or the cohort of students (the activities, culture, and norms of the course itself will create common resources between students in the same cohort). In this study, we explore the common resources of cohorts of students. We manipulate elements of the structure of course activities and observe the impact on students' behaviors, from which we infer their epistemic frames or task orientations.

The course structure manipulations are that of traditional laboratory instruction and SQILabs instruction. The two forms of instruction differ fundamentally in the way they foster scientific and critical thinking. In the SQILabs courses, significant emphasis is placed on making decisions from data \cite{HolmesPNAS}. The traditional instruction instead focuses on carrying out particular procedures to achieve a desired outcome, namely to verify physical ideas from class \cite{ALR}. An important evaluation, therefore, is whether students' behaviors differ under these two course structures. Do students' behave as through rote procedural frames or sensemaking ones?

To differentiate sensemaking from procedure-following, we consider whether the students' knowledge is active or inert. A students' knowledge may be inert when they are able to reproduce strategies or behaviors, but cannot employ them effectively on their own \cite{Bakker2011}. Related to lab activities such as data analysis, this knowledge may be made inert due to representational rather than inferential instruction \cite{Bakker2011}. Representational approaches to instruction focus on analysis concepts or tools for the purpose of representing or describing data. Inferential approaches to instruction focus on concepts or tools for the purpose of making inferences from data. Representational tools frame analysis as a final, concluding act, while inferential tools frame analysis as a stepping stone towards making sense of data. An individual tool may used in either mode. From this perspective, we can infer students' epistemological frames (sensemaking or rote procedures) from their behaviors (representing data and drawing a final conclusion or using data and analysis to make decisions for how to proceed). 

Rote procedure frames may, however, be elicited by more subtle triggers than a full semester worth of instruction. It has been found that subtle cueing to engage in formal problem solving procedures at the beginning of a task can significantly alter students' problem solving actions in the task and, ultimately, decrease their success \cite{Heckler2010, Kuo2015}. This was attributed to the fact that the cue promoted procedural or rote reasoning frames \cite{Redish2014}, rather than sensemaking frames. %

Using these two manipulations (prior instruction and cueing), we explored whether and how laboratory instruction affects students' epistemological orientation, as inferred from their experimentation behaviors. We observed students' behaviors on a common physics experiment that involved constraints of time and common measurement issues (systematic effects and mistakes). These constraints allowed us to infer their choice of orientation from their behaviors and other decision making. We study these behaviors and orientations in the context of three groups of students who differed in their prior instruction and cueing at the start of the activity under study. In the following section, we elaborate on the three conditions used, describe the experimental activity observed, and describe the ways in which we measured students' behaviors.

\section{Methods}

We compared student behavior on an identical lab experiment that provided opportunities for students to use various resources or to interpret the task in a variety of ways. The lab activity studied was deliberately selected for a number of key features. First, the activity involved a common measurement mistake and a common systematic effect (over half of the students came across these issues). Second, the activity was relatively short, so students had sufficient time to check their work, make sense of issues, perform new investigations to answer questions, or to try to improve the precision of their measurements. Third, making sense of the discrepancies would be facilitated by employing a number of data analysis tools, especially to compare measurements with uncertainty. Students' epistemological frames (especially sensemaking versus rote procedures) can be, therefore, related to whether, when, and how they decide to use various tools in their toolbox.

\subsection{Participants and conditions}

Participants were students enrolled in an introductory physics lab course at the University of British Columbia, a large, research-intensive university in Canada, across three consecutive academic years (cohorts). The course instructor, the course learning goals, and most of the physical experiment set-ups were the same across all three years. The students were also taught a similar set of data analysis tools in the same ways. Typically, instruction about a data analysis tool took place at the start of a lab session and students would be told (or expected) to apply the tool during the analysis of the experiment that followed. 

There were three conditions in the study, each a different cohort (Table \ref{tab:Conditions}): two SQILabs groups and a control group. Distinctions between the three groups, summarized in Table \ref{tab:Conditions}, focus on their prior instruction and cueing at the start of the lab activity analyzed in this study. The instructions given to the three groups about the lab experiment were identical and can be found in Appendix \ref{lab}. The instructor and teaching assistants (TAs) were instructed to only provide low-level technical support, such as with computer issues, and no behavioral support, such as to encourage students to make comparisons or improve their measurements.

\begin{table}
\centering
\begin{tabular}{C{2cm} c C{1cm} C{2cm}C{2.5cm}C{2cm}}
\hline
\hline
Group & N & Cohort Year&  Prior instruction to revise and iterate & Weighted average activity \\
\hline
Control & 136 & 1 &  None & Yes \\
SQILabs 1 & 145 & 2 & Faded scaffolding & Yes \\
SQILabs 2 & 102 & 3  & Faded scaffolding & No \\
\hline
\hline
\end{tabular}
\caption{Summary of distinctive differences between the three conditions.}
\label{tab:Conditions}
\end{table}

\begin{figure}[]
\includegraphics[width=0.3\textwidth]{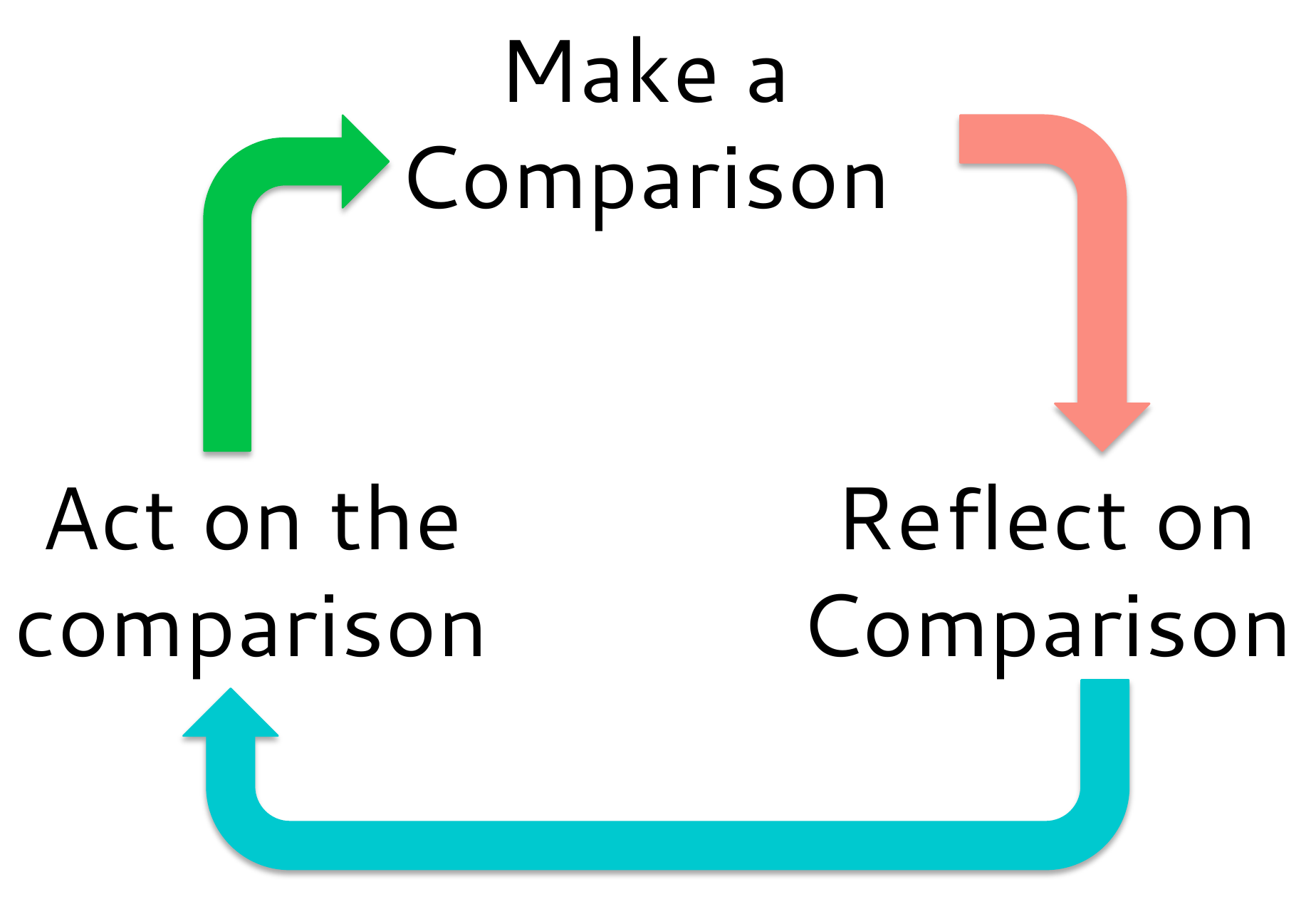}
\caption{Cycles of comparisons, reflection, and iteration in the SQILabs pedagogy from \cite{HolmesPNAS}.}\label{fig:Cycles}
\end{figure}

Prior instruction came in one of two forms: SQILabs and a control. The SQILabs curriculum focuses on sensemaking and critical thinking, as discussed in the Introduction. This involves instructing students to conduct experiments reflectively and iteratively. The basic elements of this process, shown in Fig. \ref{fig:Cycles}, involve asking students to make explicit comparisons between data (and models), reflect on and interpret those comparisons, and devise ways to act on that interpretation, especially through improving data, methods, or models \cite[see][]{HolmesPNAS, HolmesTPT,  HolmesPhD}.

This instruction was scaffolded and faded over time. In early experiments, students received explicit instructions to make particular comparisons between their data, to reflect on and interpret the result of the comparisons, and to decide how to act on the comparisons.
While iteration could involve a number of decision choices, early guided experiments focused iteration on improving the quality of measurements, especially to reduce uncertainty. This was deliberately done to confront students' assumptions that their data are inherently low quality \cite{Buffler2009, Allie1998, Sere2001}. By repeating and improving their experimental methods, students were given the opportunity to reflect on their outcomes, make decisions about how to improve their measurements, and then act on those ideas \cite{HolmesPNAS, HolmesPERC15, HolmesTPT}. The statistical and data analysis tools were critical in supporting that reflection and decision making, especially in the form of comparisons. These explicit instructions to compare, reflect, and act were slowly faded across the course.

Students in the control group conducted lab activities with the same experimental goals and physics content areas as the SQILabs groups, but were never instructed to iterate or revise their measurements. In each lab, they were asked to collect and analyze a data set to draw conclusions, usually about a given model. They were also taught the same set of data analysis tools as the SQILabs groups in the same way. The lab protocols given to the SQILabs groups once critical thinking support had been substantially faded were identical to those given to the Control group.

The two SQILabs groups differed on task cueing at the start of the activity. In the first SQILabs group (SQILabs 1), students worked through an activity about a new data analysis tool: combining uncertain measurements into a weighted average. The Control group also worked through this activity during this lab. This activity was moved to an earlier session for the SQILabs 2 group. The activity took approximately 30 minutes of student time out of a three-hour lab period. While this would increase the available time on task for the SQILabs 2 group, students had more than enough time to complete the activity in less than two and a half hours, and many students in all groups left as early as two hours into the lab period.

Both the content and the presence of the task may impact the groups' behaviors. The content of the task, combining measurements into a weighted average, is a tool primarily for representing the data in the experiment. This representational focus, as opposed to inferential, may lead students out of sensemaking frames, as discussed earlier. The presence of the task depends on the procedures associated with such tasks during prior instruction. In previous labs, it was common practice that the new data analysis tool should be employed in the analysis of the experiment that followed. The presence of the task, therefore, acts as a procedural cue and impedes sensemaking. 

From these conditions, we tested the effects of prior instruction (SQILabs versus control) and the effects of cueing with a new analysis tool. If the weighted average activity had no cueing effect on the students' subsequent actions, then the SQILabs 1 and SQILabs 2 groups should perform equally well on the outcome measures. The prior SQILabs instruction should improve whether and how students use their data analysis tools, such as to make inferences about their data or make changes to their methods to resolve discrepancies between data, if they decide to use them. These predictions relate to hypotheses that the prior SQILabs instruction creates a classroom culture of engaging heuristically and iteratively in experimentation, but that a procedural cue can impede this culture by affecting students' task orientation.

\subsection{Data collected}

We collected students' written notes from their submitted lab books to infer their experimentation behaviors. All three groups recorded their experimental process in a lab notebook that they submitted at the end of the lab period. Though students were working in pairs, each student submitted their own notebook. All groups had been instructed throughout the year to use their lab books like a journal, where they recorded everything that they did and why they were doing it. They were told to cross out things they wanted to change, rather than erasing or using `white out'. The nature of the SQILabs intervention, however, placed increased focus on recording and justifying changes and revisions in their notebooks. 

We analyzed student notes from a common experimental activity in class, the Index of Refraction lab (see Appendix \ref{lab} for the full instructions). Students were asked to use Snell's Law, Total Internal Reflection, and Brewster's angle measurements to determine the index of refraction of a piece of plexiglass ($n=1.48\pm0.02$).

This activity had three key features that facilitated our research questions. First, students commonly made a mistake when measuring the angle of refraction for the Snell's Law measurement. This mistake was a result of misreading the protractor on the equipment, such that they measured the refracted beam relative to the angle of incidence instead of relative to the normal. This mistake produced an index of refraction value around $n \approx 2$. Second, students commonly encountered a systematic effect when measuring the critical angle for Total Internal Reflection. This effect was a result of spreading of the refracted beam as the incident beam approached the critical angle. This spreading caused students to systematically overestimate the critical angle, as they continued to increase the incident angle until the entire remnants of the refracted beam had disappeared. Instead, the critical angle should have been the one at which the center of the refracted beam was about to disappear. This systematic effect shifted their index of refraction value low, typically around $n \approx 1.4$. Third, the Brewster's angle measurement was measured accurately, albeit less precisely than the other measurements. These three features created a rich environment for sensemaking.

We extracted three specific behaviors and activities from students' books: the types of analyses they performed, the follow-up from that analysis, and whether they made changes to their measurements (independent of analyses).

\subsubsection{Analyses performed by students}

We evaluate two types of analyses performed by the students: whether and how they made comparisons between their measurements and whether they combined their measurements into a weighted average. We extracted from each lab book evidence of these explicit behaviors. 

Students compared their measurements in one of two ways. A `Relative Difference' comparison was one that commented on the absolute or relative difference between the values. For example, a student wrote in their book:
\begin{quotation}
\noindent ``We noticed that the critical angle for internal reflection yielded an `n' value much lower than the others." 
 \end{quotation}
 \noindent An `Uncertainty Difference' comparison was one that commented on whether the ranges of measurements with their uncertainties overlapped or calculated how different the values were in units of uncertainty. For example, a student wrote in their books: 
 \begin{quotation}
\noindent ``The ranges for Snell's and Brewster's agree, giving an overlap of $1.47 \pm (1.57\times10^{-5})$, but the n value from critical angle does not agree." 
 \end{quotation}

Each student, therefore, obtained two scores for the analyses performed: a 1 or 0 to indicate whether they calculated a weighted average, and a 1 or 0 to indicate whether they made a comparison. They also received a categorical score for the type of comparison they made (relative difference or uncertainty).

\subsubsection{Follow-up from comparisons}

Next we evaluated whether students used their analyses to make sense of and interpret their data, beyond simply making the comparison. We did not look at student follow-up from the weighted average analysis due its representational, rather than inferential, nature.

We extracted from each lab book evidence of the following behaviors: making comparisons (as in the previous analysis), interpreting their comparisons, proposing something new from the comparison, and/or executing or evaluating that proposal. Each behavior builds on the next, so that a student proposing something new must have made and interpreted a comparison. The four behaviors were therefore coded as levels from one to four. This process is loosely based on the modified Bloom's taxonomy \cite{Anderson1994} and described in more detail in previous work \cite{HolmesPNAS, HolmesPhD}. We provide two examples below to illustrate this coding.

One student found that their Snell's Law and Total Internal Reflection measurements differed by 2.89 units of uncertainty (2.89$\sigma$). They made the following interpretation and proposal: 

\begin{quotation}
\noindent ``This suggests they are not the same value. If we had more time we would remeasure an angle for both Snell's and the critical angle." 
\end{quotation}
\noindent Because they did not act on this proposal, this student was coded as level three, proposing something new. Note that although this group appears to have run out of time, they had already made a number of revisions to their methods, descriptions, and calculations. 

Another group calculated the difference in units of uncertainty between all three measurements and interpreted that their Snell's Law measurement was ``very off" so they proposed to 
\begin{quotation}
\noindent ``Re-analyze procedure. Error found. Should have been measured from 60$\degree$ (not apparatus' 0$\degree$)." \end{quotation}
\noindent They carried out this change and then evaluated the result: ``$n\approx1.4 \rightarrow$ much better." This student was coded as level four, evaluating the proposal.

\subsubsection{Whether students made changes}

A more coarse evaluation of students' reasoning behaviors is to evaluate whether students made changes between their measurements at all. This differs from the previous analysis by including changes that did not follow logically from explicit comparisons. For example, many students simply crossed out their recorded values in the Snell's Law measurement to correct the common mistake. From the books, we cannot assess the reasoning that led to this behavior, but this analysis allows us to include this change as a productive reflection on their data (because something clearly motivated this correction). Students could receive a score of 1 or 0 separately for Snell's law and Total Internal Reflection to indicate whether they changed those measurements. We include only students who had made initially inaccurate measurements in each case. 

The accuracy of students' measurements were determined from cut-offs derived from the distribution of students' measurements and measurements made by the teaching staff. For example, an accurate Snell's Law measurement was one that fell within a 10-degree range around the expert value, the size of this range determined by the width of the peak around the expert value (the distribution of students' measured values was bimodal in this case). 

We should note that this analysis may be more accurately referred to as whether students \emph{recorded} changes, as they may well have made changes in their methods or reasoning throughout the lab that were not recorded in their books.

\subsection{Analysis methods}

All analyses were performed in R \cite{R} (see Section \ref{sec:Analysis} for package details). Here we justify the two types of statistical analyses used by evaluating the assumptions of each test. 

\subsubsection{Repeated-measures logistic regression}
We used repeated-measures logistic regression for the tests of dichotomous data (whether students chose each analysis tool and whether students made changes to each measurement). 

Our data meet the first assumption that the dependent variable is binary and represents an event occurring or not. In our sample, our dependent variables are whether students used a particular tool and whether students made changes in the two analyses. Regular logistic regression requires that each observation be independent. In our analyses, students appear twice in our sample (they could have made comparisons and/or calculated weighted averages, and they could have made changes to Snell's Law and/or Total Internal Reflection). Repeated measures logistic regression was, therefore, used instead. Logistic regression requires little or no multicollinearity, such as that each independent variable is independent from the others. In our case, the student group is always independent of the other independent variable (analysis tool or measurement type). Logistic regression also assumes linearity of the independent variables and the log odds. We meet this assumption because our independent variables are all categorical (group and analysis performed or measurement to be changed). Finally, it requires large sample sizes, which we meet with over 100 students in each group. Logistic regression also requires appropriate fitting, with only meaningful variables included. Whenever possible, alternative fitting models were considered and evaluated based on Akaike Information Criterion (AIC) and Bayesian Information Criterion (BIC).

\subsubsection{Ordinal logistic regression}
We used ordinal logistic regression for the test of how students followed up on comparisons. This analysis has similar assumptions to binary logistic regression (above). The dependent variable is an ordinal categorical variable with four ordered levels, each of which represents an event occurring (students either made explicit comparisons, interpreted those comparisons, made a proposal about the comparisons, or executed or evaluated the proposal). The observations in this case are independent such that each participant only appears once in the data set. The independent variables are also independent (no collinearity between group and follow-up behavior). The independent variables are also categorical (or ordinal) and so satisfy the assumption of linearity with the log odds. Finally we satisfy the need for large sample sizes again.

\section{Results}

We present results based on the three sets of analyses performed on students' written lab book notes. 

\subsection{Analysis tools used}

\begin{figure}[htbp]
\includegraphics[width=0.5\textwidth]{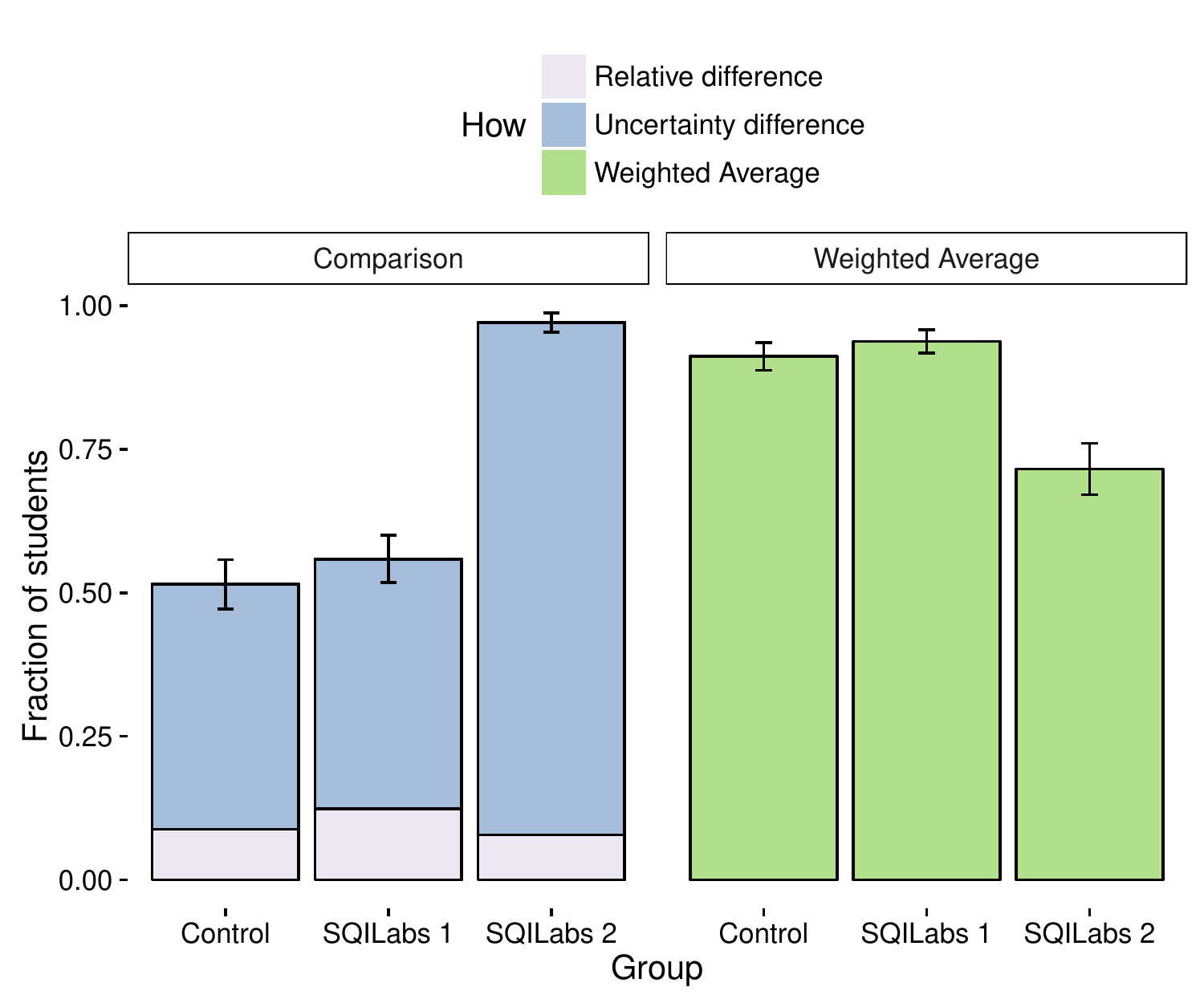}
\caption{The distribution of whether and how students analyzed their data, either using a weighted average to combine measurements or to compare their measurements. Students compared measurements either by comparing the relative difference between the measurements or the difference relative to the uncertainty.}
\label{fig:Tools}
\end{figure}

Figure \ref{fig:Tools} shows the percentage of students in each group who made explicit comparisons between their measurements and/or calculated a weighted average. Students either made comparisons by looking at the absolute or relative difference between their measurements (for example, $n=2$ is much bigger than $n=1.47$) or by comparing the difference relative to the uncertainty (for example, by calculating the ratio of the difference over the uncertainty or looking at whether the ranges of the uncertainties overlap).

There were no significant differences between the analysis choices of the control or SQILabs 1 group (see Table \ref{tab:ToolsRegress} in the Supplementary Materials). There was a statistically significant decrease in the fraction of SQILabs 2 students calculating weighted averages (compared to both the control and SQILabs 1 groups) and a statistically significant increase in the fraction of SQILabs 2 students making explicit comparisons. There was, surprisingly, no substantial change in the fraction of students making relative difference comparisons (that is, without uncertainty) across the three groups. 

These results support the hypothesis that the weighted average activity cue impacted students' sensemaking. Further interpretion will be facilitated by the other results.

\subsection{Follow-up on comparisons}

\begin{figure}[htbp]
\includegraphics[width=0.5\textwidth]{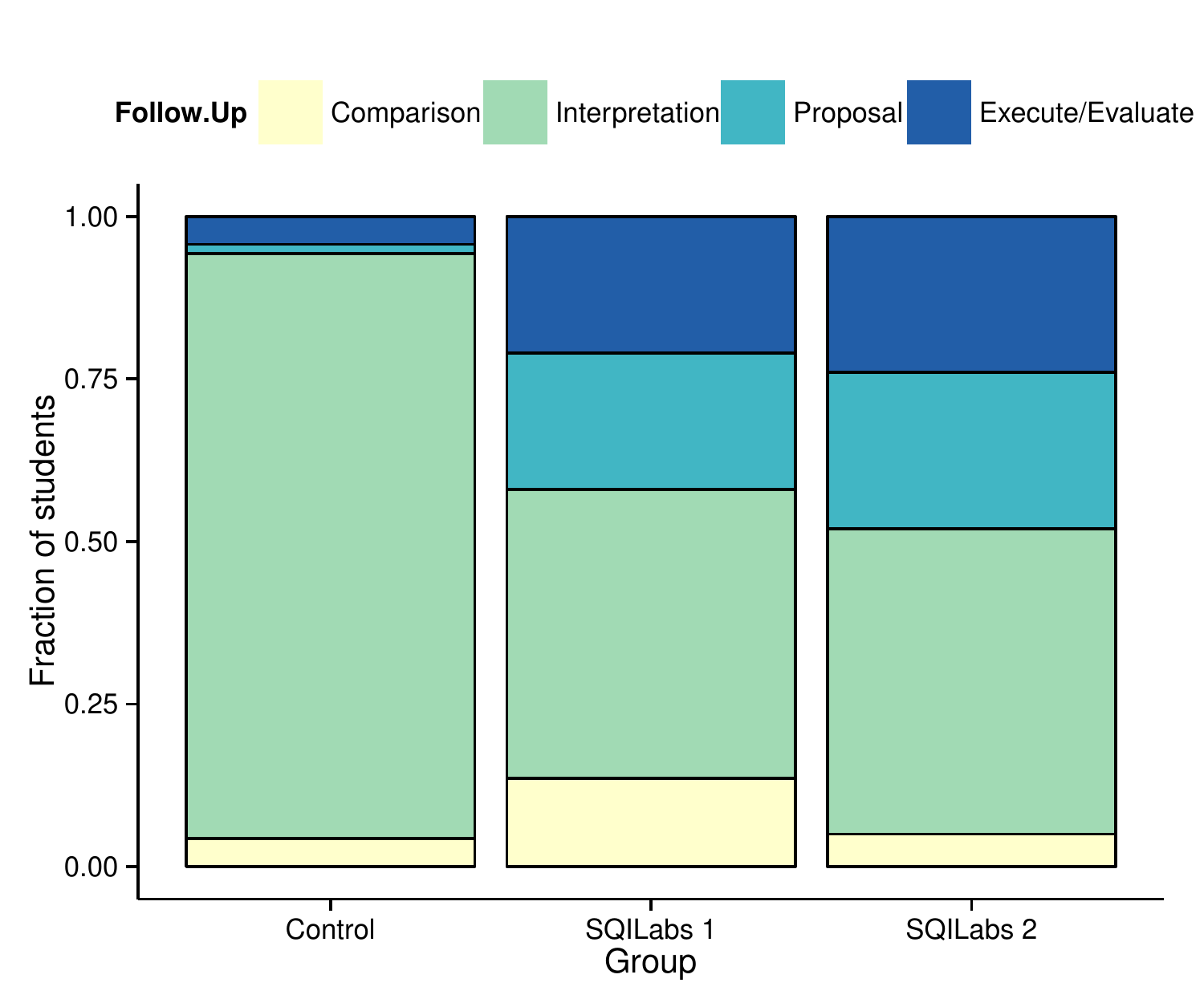}
\caption{The distribution of how students followed-up on making comparisons (as a fraction of students who made comparisons). Each level builds on the next, such that the fraction of students shaded as `Comparison' only made comparisons, students shaded as `Interpretation' both made and interpreted their comparisons and so on.}
\label{fig:FollowUp}
\end{figure}

Figure \ref{fig:FollowUp} shows the distribution of how students acted on their comparisons as a fraction of students who made comparisons. When using the control as the base group, there was a statistically significant increase in the quality of students' follow-up behaviors by both SQILabs groups. When using the SQILabs 1 group as the base, we see no statistical difference between SQILabs 1 to SQILabs 2 (See Table \ref{tab:FURegress} in the Supplementary Materials for the results of the regression analyses).

This supports the hypothesis that the prior instruction by the SQILabs improved the quality of students' sensemaking and their decision making when students made comparisons. These results also suggest that the cue does not impact the quality of students' sensemaking.
	
\subsection{Whether students made changes to their measurements}

To evaluate whether students made changes, we first need to reflect on whether students needed to make changes. From Table \ref{tab:Accuracy}, we see that over half of the class had initially inaccurate measurements for both Snell's Law and Total Internal Reflection. Accurate measurements were defined based on the accurate measurements made by the course instructor and graduate teaching assistants. More liberal ranges for the accurate values were based on what would produce reasonable agreement between the measured and expert index of refraction values (such as within one or two units of uncertainty). For Snell's Law, the accurate instructor value was 35.7$\degree \pm 0.2\degree$. The accepted range of accurate values was, therefore, between 30$\degree$ and 40$\degree$. For Total Internal Reflection, the accurate instructor value was 42.5$\degree \pm 0.5\degree$. The accepted range of accurate values was, therefore, between 40$\degree$ and 43.5$\degree$ \footnote{For those readers surprised by the narrowness of the Total Internal Reflection range compared with Snell's Law, we highlight here the non-linearity between the angle and the calculated index of refraction.}.

Table \ref{tab:Accuracy} also includes the fraction of students who reported final accurate measurements. The shifts between initial and final accuracy begin to demonstrate how many (or how few) students changed their measurements. From the results, we see that many more students in SQILabs 1 made initially inaccurate measurements. It is unclear why this is. To more critically evaluate these shifts, we examined only the students who reported initially inaccurate measurements.

\begin{table}
\centering
\caption{Percentage of students in each group with accurate Snell's Law and Total Internal Reflection measurements initially and after any changes (finally).}
\begin{tabular}{lcc}
\hline
\hline
Measurement & Group & \multicolumn{1}{p{4cm}}{Students with accurate value (Initial $\rightarrow$ Final)}\\
\hline
\multirow{3}{*}{Snell's Law} & Control & 40\% $\rightarrow$ 79\%\\
& SQILabs 1 & 28\% $\rightarrow$ 88\% \\
& SQILabs 2 & 40\% $\rightarrow$ 88\% \\
\hline
\multirow{3}{2cm}{Total Internal Reflection} & Control & 33\% $\rightarrow$ 34\% \\
& SQILabs 1 & 17\% $\rightarrow$ 21\% \\
& SQILabs 2 & 26\% $\rightarrow$ 44\% \\
\hline
\hline
\end{tabular}
\label{tab:Accuracy}
\end{table}%

\begin{figure}[htbp]
\includegraphics[width=0.5\textwidth]{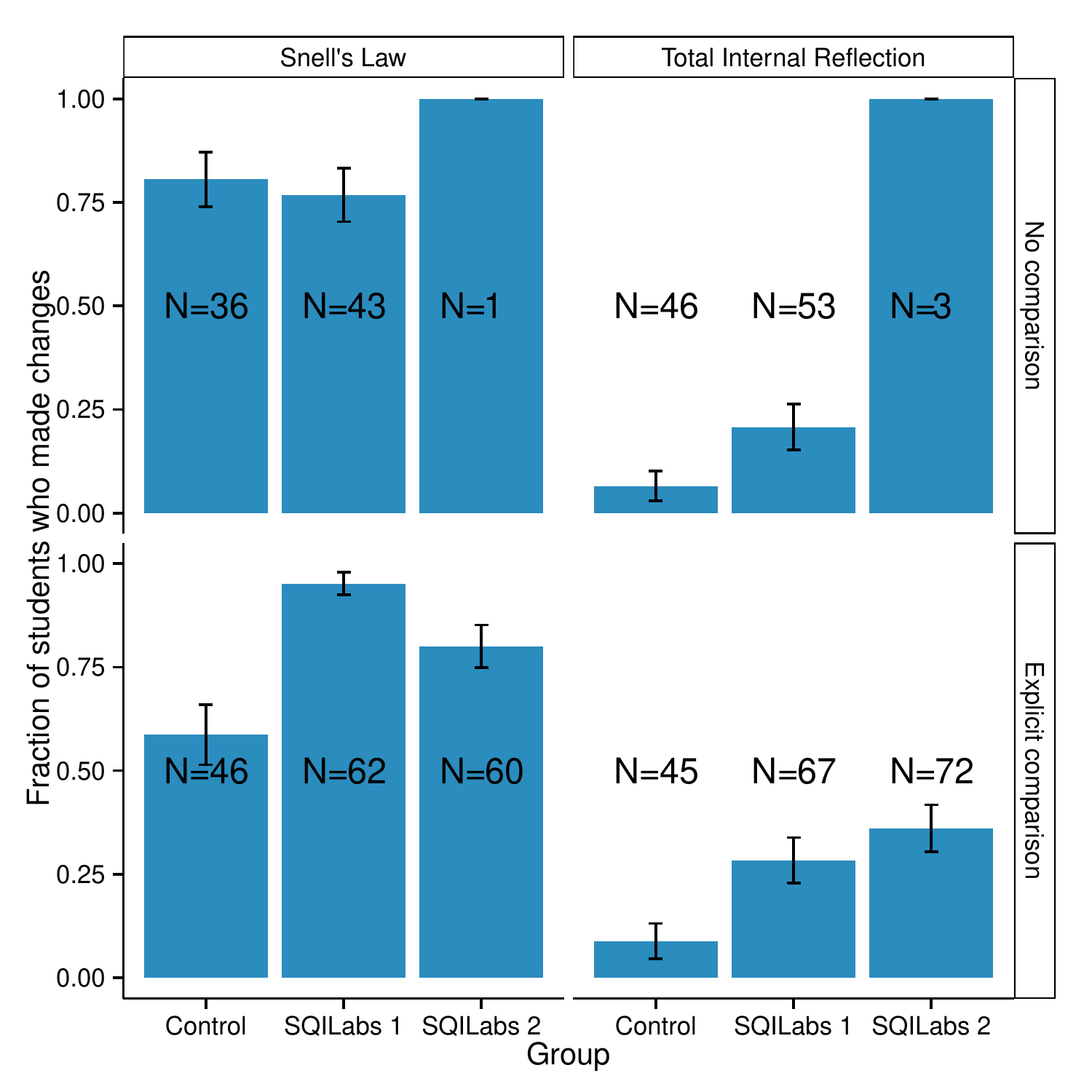}
\caption{The figure shows the fraction of students who made changes to each of their measurements, including only students who had initially inaccurate measurements, and separated by whether they made explicit comparisons between their measurements. The ``N" labels are the number of students who did or did not make comparisons in each corresponding bin, so the height of the bars represent the number students who made changes as a fraction of those who did (bottom row) or did not (top row) make comparisons and had initially inaccurate measurements.}
\label{fig:Changes}
\end{figure}

Figure \ref{fig:Changes} shows the fraction of students who made changes to their Snell's Law and Total Internal Reflection measurements, including only students with initially inaccurate measurements in each case. The results are divided into whether they had made explicit comparisons between their measurements. The ``N" labels represent the number of students with initially inaccurate measurements for each measurement that did or did not make changes to the measurement. The fractions in the figure, therefore, represent the ratio of students who made changes as a fraction of those with initially inaccurate measurements (for each measurement) and who did (or did not) make comparisons. For example, 36 students in the control group did not make comparisons and initially reported an inaccurate Snell's Law measurement. Just over 75\% of them made changes to their Snell's Law measurement. In contrast, 46 students in the control group did not make comparisons and initially reported an inaccurate Total Internal Reflection measurement. Just over 5\% of them made changes to their Total Internal Reflection measurement.

Due to convergence issues performing a three-way interaction regression, we opted to analyze the groups of students who did and did not make comparisons separately. That is, we first compared groups on whether students made changes to Snell's Law using only the students who did not make comparisons (See Table \ref{tab:NoCompChangeRegress} in the Supplementary Materials). We cannot compare students who did not make comparisons for the Total Internal Reflection measurement due to the low number of students in this category. We then separately evaluated, of the students who did make comparisons, the difference between groups on whether they made changes to either measurement (see Table \ref{tab:CompChangeRegress} in the Supplementary Materials). See Appendix \ref{sec:ChangeIssues} in the Supplementary Materials for details.

First, we examine the changes made by students who did not make comparisons. There was no significant difference between groups in the fraction of students who made changes to their Snell's Law measurement (Fig. \ref{fig:Changes} and Table \ref{tab:NoCompChangeRegress} in the Supplementary Materials). There were too few students who revised the Total Internal Reflection measurement to draw any conclusions. The small numbers of students who did not make comparisons in the SQILabs 2 group means we can only make sense of the differences between the Control and SQILabs 1 groups. All that can be said, therefore, is that prior instruction does not affect change making if students do not make comparisons.

Next, we evaluate the changes made by students who did make comparisons (second row of Fig. \ref{fig:Changes}). There were statistically significant differences between measurements and between groups, such that the SQILabs groups both outperformed the Control group for both Snell's Law and Total Internal Reflection (see Table \ref{tab:CompChangeRegress} in the Supplementary Materials). There are no statistically significant differences, however, between the two SQILabs groups. This suggests that prior instruction, and not cueing, is particularly responsible for whether students make changes to their measurements.

The results also demonstrate that more students changed their Snell's Law measurement than their Total Internal Reflection measurement across groups. While this does not directly answer our research questions, it is an interesting result that deserves future investigation (and a quick note here). We hypothesize that this difference is due to the nature of the measurement issues in each case. For Snell's law, the inaccurate measurements were primarily due to a measurement mistake, which could be quickly resolved by correcting ones calculations. For Total Internal Reflection, the inaccurate measurements were due to a systematic effect, requiring an evaluation of the model with which they were making their measurements (that is, evaluating what constitutes the critical angle at which the refracted beam is invisible). Prior work has suggested that students are more likely to correct mistakes than to make other improvements to measurements \cite{Sere1993, Kanari2004}. Prior work evaluating SQILabs instruction, specifically, has shown that students are better prepared to evaluate assumptions of a model based on evidence \cite{HolmesPNAS}, suggesting the need for an alternative hypothesis. We propose, therefore, that the relative sizes of the discrepancies is important for understanding these results. The Snell's Law mistake produced an index of refraction around $n=2$, while the Total Internal Reflection systematic effect produced an index of refraction around $n=1.41$. The Brewster's angle measurement accurately produced a value around $n=1.47$, with no common mistakes or systematics. It may have been much more apparent to students that $n=2$ was the `odd one out,' while the $n=1.41$ difference was relatively subtle. These hypotheses make way for more detailed evaluation of the mechanisms through which revisions to measurements are made in these courses.

\section{Discussion}

In this study, we evaluated the effects of prior instruction and activity structure on how students engaged with data analysis tools and experimentation behaviors. The results suggest a shift in students' orientation towards the lab activity due to both the prior instruction and activity structure. The SQILabs course structure, compared to the Control group, did not significantly affect whether they used particular data analysis tools. It did, however, significantly improve how students acted on their data analysis, with higher-level follow-up on their results and more changes to improve their measurements. These results are consistent with prior work \cite{HolmesPNAS}.

The presence of the priming task, on the other hand, did affect whether students used particular data analysis tools. When the lab activity began with instruction about a new data analysis tool (a weighted average), more students calculated a weighted average and fewer made comparisons between their three nominally equivalent measurements. This is particularly noteworthy because making comparisons between measurements was a more useful analysis tool for thinking critically about their results and experiment. 

We suggest from these results a set of related explanations. First, we highlight that the SQILabs 2 group were equipped with a full data analysis toolbox and were not learning any new tools that day. We propose that students in this group had the opportunity to decide which tool (or tools) to employ. This independent decision making allowed them to select the tools that more appropriately supported their critical thinking skills, provided by the prior SQILabs instruction. In contrast, the Control and SQILabs 1 group had both learned a new tool, and so they assumed (as had been true in previous lab activities) that they needed to apply their new tool. This application came at the expense of considering other tools --- that is, thinking critically about which tools were relevant to make sense of their data.

A related perspective comes from the pair of studies on cueing with formal procedures, described in the introduction \cite{Heckler2010, Kuo2015}. The authors in these studies showed that when students were given cues to follow a formal procedure to solve a problem, they were less likely to obtain the correct answer than students who were not cued. It was found that the formal cue decreased the likelihood that students would use conceptual shortcuts or intuition that were productive to successfully solving the problem. They suggested that the formal procedures may have suppressed students' flexible reasoning, potentially leading to less adaptive problem solving. As described in \cite[][p. 1848]{Heckler2010}, the prompts ``may tend to cue undesired epistemological resources in novice students."

These two explanations (the formal procedure cueing and the heuristic decisions from a full toolbox) are further evidenced by students' follow-up behaviors. The SQILabs 1 students who chose to make explicit comparisons engaged in similarly high-level follow-up behaviors as the SQILabs 2 group (and both groups outperformed the Control group). When the SQILabs 1 students did not make comparisons, they performed just as well as the Control group. This result was also seen in the Kuo, et al. study \cite{Kuo2015}, where the students who used a conceptual solution approach were more likely to obtain a correct answer than those who did not, regardless of whether they were primed to use the formal approach. While the cue may shift students to apply the formal approach, the best outcomes are when students decide on their own which tool or approach to use.

These explanations would have been strengthened by having a Control group that did not receive the formal cue at the start of the activity. Indeed the practicalities and ethics of conducting research in situ place limitations on the availability of conditions. Further research should also explore, qualitatively, the mechanisms through which students make changes to further understand this process. That is, while students' written documents provide significant insight into students' process, video recordings of their actions would provide substantial evidence of what does and does not trigger sensemaking in the course.

This research has, once again, demonstrated how cueing can result in student behaviors that are consistent with those of students engaged in procedural, rather than sensemaking frames. It is less clear whether behavioral cues can also trigger sensemaking frames over procedural ones. That is, would a cue to reflect and iterate during their experiment have promoted sensemaking or would it transform these otherwise effective strategies into procedural ones?

\section{Conclusions}
This study explored the impact of procedural cues and prior instruction on students' heuristic behaviors in conducting a lab experiment. Prior instruction on conducting experiments reflectively and iteratively was shown to improve students' reasoning and thinking with their analysis tools, extending existing work on the efficacies of the SQILabs curriculum for introductory physics labs. The same prior instruction with a procedural cue, however, hindered the likelihood that students would choose an appropriate analysis tool, and instead focused on following the formal procedures from the cue. If students did break from the formal cue and applied an appropriate analysis tool, the prior instruction supported them in engaging in high-level follow-up reasoning and revising inaccurate measurements.

The subtlety of the cueing effect has implications for general instruction. From the results here, it is clear that instructors need to pay careful attention to what might be cued inadvertently. An extra task intended to contribute to or complement an activity can cue undesirable epistemological frames and behaviors. Instructors may also need to explicitly address students' epistemological frames or task orientations to counteract the effects of formal procedure cues, when necessary. One strategy that may help to achieve this, introduction of new tools or procedural tasks ought to be spaced over time, such that students have opportunities to practice selecting tools and procedures from an increasingly full toolbox, without the introduction of a shiny new tool.

\bibliography{Bib.bib}

\begin{thebibliography}{24}%
\makeatletter
\providecommand \@ifxundefined [1]{%
 \@ifx{#1\undefined}
}%
\providecommand \@ifnum [1]{%
 \ifnum #1\expandafter \@firstoftwo
 \else \expandafter \@secondoftwo
 \fi
}%
\providecommand \@ifx [1]{%
 \ifx #1\expandafter \@firstoftwo
 \else \expandafter \@secondoftwo
 \fi
}%
\providecommand \natexlab [1]{#1}%
\providecommand \enquote  [1]{``#1''}%
\providecommand \bibnamefont  [1]{#1}%
\providecommand \bibfnamefont [1]{#1}%
\providecommand \citenamefont [1]{#1}%
\providecommand \href@noop [0]{\@secondoftwo}%
\providecommand \href [0]{\begingroup \@sanitize@url \@href}%
\providecommand \@href[1]{\@@startlink{#1}\@@href}%
\providecommand \@@href[1]{\endgroup#1\@@endlink}%
\providecommand \@sanitize@url [0]{\catcode `\\12\catcode `\$12\catcode
  `\&12\catcode `\#12\catcode `\^12\catcode `\_12\catcode `\%12\relax}%
\providecommand \@@startlink[1]{}%
\providecommand \@@endlink[0]{}%
\providecommand \url  [0]{\begingroup\@sanitize@url \@url }%
\providecommand \@url [1]{\endgroup\@href {#1}{\urlprefix }}%
\providecommand \urlprefix  [0]{URL }%
\providecommand \Eprint [0]{\href }%
\providecommand \doibase [0]{http://dx.doi.org/}%
\providecommand \selectlanguage [0]{\@gobble}%
\providecommand \bibinfo  [0]{\@secondoftwo}%
\providecommand \bibfield  [0]{\@secondoftwo}%
\providecommand \translation [1]{[#1]}%
\providecommand \BibitemOpen [0]{}%
\providecommand \bibitemStop [0]{}%
\providecommand \bibitemNoStop [0]{.\EOS\space}%
\providecommand \EOS [0]{\spacefactor3000\relax}%
\providecommand \BibitemShut  [1]{\csname bibitem#1\endcsname}%
\let\auto@bib@innerbib\@empty
\bibitem [{\citenamefont {Holmes}\ \emph {et~al.}(2015)\citenamefont {Holmes},
  \citenamefont {Wieman},\ and\ \citenamefont {Bonn}}]{HolmesPNAS}%
  \BibitemOpen
  \bibfield  {author} {\bibinfo {author} {\bibfnamefont {N.}~\bibnamefont
  {Holmes}}, \bibinfo {author} {\bibfnamefont {C.~E.}\ \bibnamefont {Wieman}},
  \ and\ \bibinfo {author} {\bibfnamefont {D.}~\bibnamefont {Bonn}},\ }\href
  {\doibase 10.1073/pnas.1505329112} {\bibfield  {journal} {\bibinfo  {journal}
  {PNAS}\ }\textbf {\bibinfo {volume} {112}},\ \bibinfo {pages} {11199}
  (\bibinfo {year} {2015})}\BibitemShut {NoStop}%
\bibitem [{\citenamefont {Holmes}\ and\ \citenamefont
  {Bonn}(2015)}]{HolmesTPT}%
  \BibitemOpen
  \bibfield  {author} {\bibinfo {author} {\bibfnamefont {N.~G.}\ \bibnamefont
  {Holmes}}\ and\ \bibinfo {author} {\bibfnamefont {D.~A.}\ \bibnamefont
  {Bonn}},\ }\href {\doibase 10.1119/1.4928350} {\bibfield  {journal} {\bibinfo
   {journal} {The Physics Teacher}\ }\textbf {\bibinfo {volume} {53}},\
  \bibinfo {pages} {352} (\bibinfo {year} {2015})}\BibitemShut {NoStop}%
\bibitem [{\citenamefont {Hofstein}\ and\ \citenamefont
  {Lunetta}(1982)}]{Hofstein1982}%
  \BibitemOpen
  \bibfield  {author} {\bibinfo {author} {\bibfnamefont {A.}~\bibnamefont
  {Hofstein}}\ and\ \bibinfo {author} {\bibfnamefont {V.~N.}\ \bibnamefont
  {Lunetta}},\ }\href {\doibase 10.3102/00346543052002201} {\bibfield
  {journal} {\bibinfo  {journal} {Review of Educational Research}\ }\textbf
  {\bibinfo {volume} {52}},\ \bibinfo {pages} {201} (\bibinfo {year}
  {1982})}\BibitemShut {NoStop}%
\bibitem [{\citenamefont {Hofstein}\ and\ \citenamefont
  {Lunetta}(2004)}]{Hofstein2004}%
  \BibitemOpen
  \bibfield  {author} {\bibinfo {author} {\bibfnamefont {A.}~\bibnamefont
  {Hofstein}}\ and\ \bibinfo {author} {\bibfnamefont {V.~N.}\ \bibnamefont
  {Lunetta}},\ }\href {\doibase 10.1002/sce.10106} {\bibfield  {journal}
  {\bibinfo  {journal} {Science Education}\ }\textbf {\bibinfo {volume} {88}},\
  \bibinfo {pages} {28} (\bibinfo {year} {2004})}\BibitemShut {NoStop}%
\bibitem [{ALR(2005)}]{ALR}%
  \BibitemOpen
  \href {http://www.nap.edu/catalog.php?record{\_}id=11311} {\emph {\bibinfo
  {title} {{America's Lab Report: Investigations in High School Science}}}},\
  \bibinfo {type} {Tech. Rep.}\ (\bibinfo  {institution} {Committee on High
  School Science Laboratories: Role and Vision; National Research Council},\
  \bibinfo {address} {Washington, D.C.},\ \bibinfo {year} {2005})\BibitemShut
  {NoStop}%
\bibitem [{\citenamefont {Bing}\ and\ \citenamefont {Redish}(2009)}]{Bing2009}%
  \BibitemOpen
  \bibfield  {author} {\bibinfo {author} {\bibfnamefont {T.~J.}\ \bibnamefont
  {Bing}}\ and\ \bibinfo {author} {\bibfnamefont {E.~F.}\ \bibnamefont
  {Redish}},\ }\href {\doibase 10.1103/PhysRevSTPER.5.020108} {\bibfield
  {journal} {\bibinfo  {journal} {Physical Review Special Topics - Physics
  Education Research}\ }\textbf {\bibinfo {volume} {5}},\ \bibinfo {pages}
  {020108} (\bibinfo {year} {2009})}\BibitemShut {NoStop}%
\bibitem [{\citenamefont {Butler}\ and\ \citenamefont
  {Cartier}(2004)}]{Butler2004}%
  \BibitemOpen
  \bibfield  {author} {\bibinfo {author} {\bibfnamefont {D.~L.}\ \bibnamefont
  {Butler}}\ and\ \bibinfo {author} {\bibfnamefont {S.~C.}\ \bibnamefont
  {Cartier}},\ }\href@noop {} {\bibfield  {journal} {\bibinfo  {journal}
  {Teachers College Record}\ }\textbf {\bibinfo {volume} {106}},\ \bibinfo
  {pages} {1729 } (\bibinfo {year} {2004})}\BibitemShut {NoStop}%
\bibitem [{\citenamefont {Bakker}\ and\ \citenamefont
  {Derry}(2011)}]{Bakker2011}%
  \BibitemOpen
  \bibfield  {author} {\bibinfo {author} {\bibfnamefont {A.}~\bibnamefont
  {Bakker}}\ and\ \bibinfo {author} {\bibfnamefont {J.}~\bibnamefont {Derry}},\
  }\href {\doibase 10.1080/10986065.2011.538293} {\bibfield  {journal}
  {\bibinfo  {journal} {Mathematical Thinking and Learning}\ }\textbf {\bibinfo
  {volume} {13}},\ \bibinfo {pages} {5} (\bibinfo {year} {2011})}\BibitemShut
  {NoStop}%
\bibitem [{\citenamefont {Heckler}(2010)}]{Heckler2010}%
  \BibitemOpen
  \bibfield  {author} {\bibinfo {author} {\bibfnamefont {A.~F.}\ \bibnamefont
  {Heckler}},\ }\href {\doibase 10.1080/09500690903199556} {\bibfield
  {journal} {\bibinfo  {journal} {International Journal of Science Education}\
  }\textbf {\bibinfo {volume} {32}},\ \bibinfo {pages} {1829} (\bibinfo {year}
  {2010})}\BibitemShut {NoStop}%
\bibitem [{\citenamefont {Kuo}\ \emph {et~al.}(2015)\citenamefont {Kuo},
  \citenamefont {Hallinen},\ and\ \citenamefont {Conlin}}]{Kuo2015}%
  \BibitemOpen
  \bibfield  {author} {\bibinfo {author} {\bibfnamefont {E.}~\bibnamefont
  {Kuo}}, \bibinfo {author} {\bibfnamefont {N.~R.}\ \bibnamefont {Hallinen}}, \
  and\ \bibinfo {author} {\bibfnamefont {L.~D.}\ \bibnamefont {Conlin}},\ }in\
  \href {\doibase 10.1119/perc.2015.pr.041} {\emph {\bibinfo {booktitle} {2015
  Physics Education Research Conference Proceedings}}}\ (\bibinfo  {publisher}
  {American Association of Physics Teachers},\ \bibinfo {year} {2015})\ pp.\
  \bibinfo {pages} {183--186}\BibitemShut {NoStop}%
\bibitem [{\citenamefont {Redish}(2014)}]{Redish2014}%
  \BibitemOpen
  \bibfield  {author} {\bibinfo {author} {\bibfnamefont {E.~F.}\ \bibnamefont
  {Redish}},\ }\href {\doibase 10.1119/1.4874260} {\bibfield  {journal}
  {\bibinfo  {journal} {American Journal of Physics}\ }\textbf {\bibinfo
  {volume} {82}},\ \bibinfo {pages} {537} (\bibinfo {year} {2014})}\BibitemShut
  {NoStop}%
\bibitem [{\citenamefont {Holmes}(2015{\natexlab{a}})}]{HolmesPhD}%
  \BibitemOpen
  \bibfield  {author} {\bibinfo {author} {\bibfnamefont {N.~G.}\ \bibnamefont
  {Holmes}},\ }\emph {\bibinfo {title} {{Structured quantitative inquiry labs:
  developing critical thinking in the introductory physics laboratory}}},\
  \href {http://circle.ubc.ca/handle/2429/51363} {Ph.D. thesis},\ \bibinfo
  {school} {University of British Columbia} (\bibinfo {year}
  {2015}{\natexlab{a}})\BibitemShut {NoStop}%
\bibitem [{\citenamefont {Buffler}\ \emph {et~al.}(2009)\citenamefont
  {Buffler}, \citenamefont {Lubben},\ and\ \citenamefont
  {Ibrahim}}]{Buffler2009}%
  \BibitemOpen
  \bibfield  {author} {\bibinfo {author} {\bibfnamefont {A.}~\bibnamefont
  {Buffler}}, \bibinfo {author} {\bibfnamefont {F.}~\bibnamefont {Lubben}}, \
  and\ \bibinfo {author} {\bibfnamefont {B.}~\bibnamefont {Ibrahim}},\ }\href
  {\doibase 10.1080/09500690802189807} {\bibfield  {journal} {\bibinfo
  {journal} {International Journal of Science Education}\ }\textbf {\bibinfo
  {volume} {31}},\ \bibinfo {pages} {1137} (\bibinfo {year}
  {2009})}\BibitemShut {NoStop}%
\bibitem [{\citenamefont {Allie}\ and\ \citenamefont
  {Buffler}(1998)}]{Allie1998}%
  \BibitemOpen
  \bibfield  {author} {\bibinfo {author} {\bibfnamefont {S.}~\bibnamefont
  {Allie}}\ and\ \bibinfo {author} {\bibfnamefont {A.}~\bibnamefont
  {Buffler}},\ }\href {\doibase 10.1119/1.18915} {\bibfield  {journal}
  {\bibinfo  {journal} {American Journal of Physics}\ }\textbf {\bibinfo
  {volume} {66}},\ \bibinfo {pages} {613} (\bibinfo {year} {1998})}\BibitemShut
  {NoStop}%
\bibitem [{\citenamefont {S{\'{e}}r{\'{e}}}\ \emph {et~al.}(2001)\citenamefont
  {S{\'{e}}r{\'{e}}}, \citenamefont {Fernandez-Gonzalez}, \citenamefont
  {Gallegos}, \citenamefont {Gonzalez-Garcia}, \citenamefont {Manuel},
  \citenamefont {Perales},\ and\ \citenamefont {Leach}}]{Sere2001}%
  \BibitemOpen
  \bibfield  {author} {\bibinfo {author} {\bibfnamefont {M.-G.}\ \bibnamefont
  {S{\'{e}}r{\'{e}}}}, \bibinfo {author} {\bibfnamefont {M.}~\bibnamefont
  {Fernandez-Gonzalez}}, \bibinfo {author} {\bibfnamefont {J.~A.}\ \bibnamefont
  {Gallegos}}, \bibinfo {author} {\bibfnamefont {F.}~\bibnamefont
  {Gonzalez-Garcia}}, \bibinfo {author} {\bibfnamefont {E.~D.}\ \bibnamefont
  {Manuel}}, \bibinfo {author} {\bibfnamefont {F.~J.}\ \bibnamefont {Perales}},
  \ and\ \bibinfo {author} {\bibfnamefont {J.}~\bibnamefont {Leach}},\ }\href
  {\doibase 10.1023/A:1013141706723} {\bibfield  {journal} {\bibinfo  {journal}
  {Research in Science Education}\ }\textbf {\bibinfo {volume} {31}},\ \bibinfo
  {pages} {499} (\bibinfo {year} {2001})}\BibitemShut {NoStop}%
\bibitem [{\citenamefont {Holmes}(2015{\natexlab{b}})}]{HolmesPERC15}%
  \BibitemOpen
  \bibfield  {author} {\bibinfo {author} {\bibfnamefont {N.}~\bibnamefont
  {Holmes}},\ }in\ \href {\doibase 10.1119/perc.2015.plenary.001} {\emph
  {\bibinfo {booktitle} {Physics Education Research Conference 2015}}},\
  \bibinfo {editor} {edited by\ \bibinfo {editor} {\bibfnamefont {A.~D.}\
  \bibnamefont {Churukian}}, \bibinfo {editor} {\bibfnamefont {D.~L.}\
  \bibnamefont {Jones}}, \ and\ \bibinfo {editor} {\bibfnamefont
  {L.}~\bibnamefont {Ding}}}\ (\bibinfo {address} {College Park, MD},\ \bibinfo
  {year} {2015})\ pp.\ \bibinfo {pages} {14--17}\BibitemShut {NoStop}%
\bibitem [{\citenamefont {Anderson}\ and\ \citenamefont
  {A.}(1994)}]{Anderson1994}%
  \BibitemOpen
  \bibfield  {author} {\bibinfo {author} {\bibfnamefont {L.}~\bibnamefont
  {Anderson}}\ and\ \bibinfo {author} {\bibfnamefont {S.~L.}\ \bibnamefont
  {A.}},\ }in\ \href@noop {} {\emph {\bibinfo {booktitle} {National Society for
  the Study of Education Yearbooks}}}\ (\bibinfo  {publisher} {University of
  Chicago Press},\ \bibinfo {year} {1994})\BibitemShut {NoStop}%
\bibitem [{\citenamefont {{R Core Team}}(2016)}]{R}%
  \BibitemOpen
  \bibfield  {author} {\bibinfo {author} {\bibnamefont {{R Core Team}}},\
  }\href {https://www.r-project.org} {\emph {\bibinfo {title} {{R: A Language
  and Environment for Statistical Computing}}}},\ \bibinfo {type} {Tech. Rep.}\
  (\bibinfo  {institution} {{R Foundation for Statistical Computing}},\
  \bibinfo {address} {Vienna, Austria},\ \bibinfo {year} {2016})\BibitemShut
  {NoStop}%
\bibitem [{Note1()}]{Note1}%
  \BibitemOpen
  \bibinfo {note} {For those readers surprised by the narrowness of the Total
  Internal Reflection range compared with Snell's Law, we highlight here the
  non-linearity between the angle and the calculated index of
  refraction.}\BibitemShut {Stop}%
\bibitem [{\citenamefont {S{\'{e}}r{\'{e}}}\ \emph {et~al.}(1993)\citenamefont
  {S{\'{e}}r{\'{e}}}, \citenamefont {Journeaux},\ and\ \citenamefont
  {Larcher}}]{Sere1993}%
  \BibitemOpen
  \bibfield  {author} {\bibinfo {author} {\bibfnamefont {M.}~\bibnamefont
  {S{\'{e}}r{\'{e}}}}, \bibinfo {author} {\bibfnamefont {R.}~\bibnamefont
  {Journeaux}}, \ and\ \bibinfo {author} {\bibfnamefont {C.}~\bibnamefont
  {Larcher}},\ }\href {\doibase 10.1080/0950069930150406} {\bibfield  {journal}
  {\bibinfo  {journal} {International Journal of Science Education}\ }\textbf
  {\bibinfo {volume} {15}},\ \bibinfo {pages} {427} (\bibinfo {year}
  {1993})}\BibitemShut {NoStop}%
\bibitem [{\citenamefont {Kanari}\ and\ \citenamefont
  {Millar}(2004)}]{Kanari2004}%
  \BibitemOpen
  \bibfield  {author} {\bibinfo {author} {\bibfnamefont {Z.}~\bibnamefont
  {Kanari}}\ and\ \bibinfo {author} {\bibfnamefont {R.}~\bibnamefont
  {Millar}},\ }\href {\doibase 10.1002/tea.20020} {\bibfield  {journal}
  {\bibinfo  {journal} {Journal of Research in Science Teaching}\ }\textbf
  {\bibinfo {volume} {41}},\ \bibinfo {pages} {748} (\bibinfo {year}
  {2004})}\BibitemShut {NoStop}%
\bibitem [{\citenamefont {Bates}\ \emph {et~al.}(2015)\citenamefont {Bates},
  \citenamefont {M{\"{a}}chler}, \citenamefont {Bolker},\ and\ \citenamefont
  {Walker}}]{lme4}%
  \BibitemOpen
  \bibfield  {author} {\bibinfo {author} {\bibfnamefont {D.}~\bibnamefont
  {Bates}}, \bibinfo {author} {\bibfnamefont {M.}~\bibnamefont
  {M{\"{a}}chler}}, \bibinfo {author} {\bibfnamefont {B.}~\bibnamefont
  {Bolker}}, \ and\ \bibinfo {author} {\bibfnamefont {S.}~\bibnamefont
  {Walker}},\ }\href {\doibase 10.18637/jss.v067.i01} {\bibfield  {journal}
  {\bibinfo  {journal} {Journal of Statistical Software}\ }\textbf {\bibinfo
  {volume} {67}},\ \bibinfo {pages} {1} (\bibinfo {year} {2015})}\BibitemShut
  {NoStop}%
\bibitem [{\citenamefont {Fox}\ and\ \citenamefont {Weisberg}(2011)}]{car}%
  \BibitemOpen
  \bibfield  {author} {\bibinfo {author} {\bibfnamefont {J.}~\bibnamefont
  {Fox}}\ and\ \bibinfo {author} {\bibfnamefont {S.}~\bibnamefont {Weisberg}},\
  }\href {http://socserv.socsci.mcmaster.ca/jfox/Books/Companion} {\emph
  {\bibinfo {title} {{An {\{}R{\}} Companion to Applied Regression}}}},\
  \bibinfo {edition} {2nd}\ ed.\ (\bibinfo  {publisher} {Sage},\ \bibinfo
  {address} {Thousand Oaks, CA},\ \bibinfo {year} {2011})\BibitemShut {NoStop}%
\bibitem [{\citenamefont {Venables}\ and\ \citenamefont {Ripley}(2002)}]{MASS}%
  \BibitemOpen
  \bibfield  {author} {\bibinfo {author} {\bibfnamefont {W.~N.}\ \bibnamefont
  {Venables}}\ and\ \bibinfo {author} {\bibfnamefont {B.~D.}\ \bibnamefont
  {Ripley}},\ }\href {{www.stats.ox.ac.uk/pub/MASS4}} {\emph {\bibinfo {title}
  {{Modern Applied Statistics with S}}}},\ \bibinfo {edition} {4th}\ ed.\
  (\bibinfo  {publisher} {Springer},\ \bibinfo {address} {New York},\ \bibinfo
  {year} {2002})\BibitemShut {NoStop}%
\end{thebibliography}%

\acknowledgements
We would like to acknowledge the extensive feedback from Carl Wieman on this manuscript, the illuminating and educational statistics discussions from Shima Salehi, and additional discussions and consultations with Sarah Gilbert, the UBC PHASER group, and the Wieman research group. We are very grateful to the reviewers of this manuscript for thoughtful and constructive feedback. 

\newpage
\begin{appendix}
\section{The Index of Refraction lab}\label{lab}

The full set of instructions provided to students in the Index of Refraction lab are as follows:

\vspace{2em}
\begin{em}
This experiment is an introduction to optics and will explore Snell's Law, total internal reflection and Brewster's angle. The experiment involves making precise optical measurements with a laser and estimating the uncertainty in those measurements. These measurements and the data analysis allow you to measure the index of refraction n of plexiglass in three different ways. 

\begin{center}\textbf{CHECK ALIGNMENT}\end{center}

Check that the optical bench is aligned by setting the incident angle in air to 0 degrees and observe the position of the final spot. Rotate the plexiglass through 180 degrees so that the incident angle through plexiglass is again 0 degrees. The position of the final spot should not have moved. If you feel that it is misaligned, contact your TA for help. DO NOT ADJUST ANY OF THE MIRRORS.

\begin{center}\textbf{SNELL'S LAW}\end{center}

When entering the flat surface, the incident beam goes from air (index of refraction ~1) to plexiglass (index of refraction is \emph{n}). By measuring the angle of the refracted beam and the angle of incidence, the index of refraction can be determined from:

\begin{equation*}n = sin(\text{incident angle})/sin(\text{refracted angle})\end{equation*}

Use an incident angle of 60 degrees for this measurement. Record your value of \emph{n} determined by this measurement, including an estimate of uncertainty.

\begin{center}\textbf{CRITICAL ANGLE FOR TOTAL INTERNAL REFLECTION}\end{center}

With the beam entering the curved surface first, search for the critical angle of incidence at which you judge total internal reflection occurs. This is the incident angle beyond which the beam is reflected, but with no refracted beam. The index of refraction can be determined from this using:

\begin{equation*}sin(\text{critical angle}) = 1/n\end{equation*}

Record your value of \emph{n} determined by this measurement, including an estimate of uncertainty.

\begin{center}\textbf{BREWSTER'S ANGLE}\end{center}

Brewster's angle is the angle of incidence at which the reflected beam is completely polarized. With the beam entering the flat surface first, determine the angle at which the reflected beam is completely polarized. At this angle, a polarizer intercepting the reflected beam can completely block it if it is oriented so that only passes the opposite polarization. The index of refraction can be determined from this using:

\begin{equation*}tan(\text{Brewter's angle}) = n\end{equation*}

Record your value of n determined by this measurement, including an estimate of uncertainty.

\begin{center}\textbf{Marking Scheme}\end{center}

High quality measurement of index of refraction 3 different ways: 12 marks

Don't forget to describe your procedures as you go along through the experiment.
\end{em}

 \newpage
 \section{Regression analyses}\label{sec:Analysis}

\vspace{-1em}We used the \emph{lme4} package in R for the repeated-measures generalized linear mixed-effects models \cite{lme4}. We used the \emph{car} package in R to obtain ANOVA tables for the statistical tests \cite{car}. We used the \emph{stats} package from the R core \cite{R} to for the generalized linear models (sub-testing in the analysis of whether students made changes). We used the MASS package in R for the order logistic regression models \cite{MASS}. We again used the \emph{car} package in R to obtain ANOVA tables for the statistical tests \cite{car}.

\vspace{-1em}\subsection{Analysis tools used}

\begin{table*}[hp]
\centering
\begin{tabular}{ccr@{.}lr@{.}lr@{.}lr@{.}lr@{.}l}
\hline
\hline
 & Comparison & \multicolumn{2}{p{1.5cm}}{Estimate (log odds)}& \multicolumn{2}{c}{Odds}& \multicolumn{2}{p{1.5cm}}{Standard Error} &\multicolumn{2}{c}{$z$-value}& \multicolumn{2}{c}{\emph{p}}\\
\hline
\multirow{6}{3cm}{Logistic Regression (Base: Control [Group]; Weighted Average [Analysis performed]} & (Intercept) & 2&34 & 10&33 & 0&30 & 7&73 & $<$&001$^{***}$\\
&Making Comparison & -2&28 & 0&10 & 0&35 & -6&55 & $<$&001$^{***}$\\
&SQILabs 1 & 0&38 & 1&46 & 0&46 & 0&83 & &407\\
&SQILabs 2 & -1&41 & 0&37 & -3&78 & 7&73 & $<$&001$^{***}$\\
&Making Comparison * SQILabs 1 & -0&20 & 0&82& 0&52 & -0&39 & &694 \\
&Making Comparison * SQILabs 2 & 4&85 & 127&74 & 0&72 & 6&78 & $<$&001$^{***}$\\
\hline\hline
 & Variable & \multicolumn{2}{c}{$\chi^2$} & \multicolumn{2}{p{1.5cm}}{Degrees of freedom} &\multicolumn{2}{c}{$p$}&&\\
\hline
& Analysis performed & 48&30 & \multicolumn{2}{c}{1} & $<$&001$^{***}$\\
ANOVA & Group & 4&00 & \multicolumn{2}{c}{2} & &136\\
& Analysis performed * Group & 53&51 & \multicolumn{2}{c}{2} & $<$&001$^{***}$\\
\hline
\hline
\end{tabular}
\caption{Logistic regression table for students' analysis tool use. Base comparison is the control group (for year) and calculating a weighted average (for analysis performed) $.$ \emph{p}$<$.1. $^{*}$ \emph{p}$<$.05. $^{**}$ \emph{p}$<$.01. $^{***}$ \emph{p}$<$.001.}
\label{tab:ToolsRegress}
\end{table*}

\vspace{-1em}Table \ref{tab:ToolsRegress} shows the results of the repeated-measures logistic regression of these comparisons, where tool use was the binary outcome variable (they either used a tool or did not) and group and analysis performed (either making comparisons or calculating a weighted average) were the input variables. The Control group was used as the base-level comparison group.

\vspace{-1em}\subsection{Follow-up on comparisons}

\vspace{-1em}Table \ref{tab:FURegress} shows the results of the ordinal logistic regression used to evaluate students' follow-up behaviors. Students' follow-up behaviors (either comparing, interpreting, proposing, or executing/evaluating comparisons) were used as an ordered categorical outcome variable because the behaviors build upon each other (a student who executed/evaluated their comparisons had to have made, interpreted, and proposed from a comparison). The input variable was group. Both the Control group and SQILabs 1 groups were used as base-level comparison groups (separately) to evaluate the differences between the two SQILabs groups.

\begin{table*}
\centering
\begin{tabular}{ccr@{.}lr@{.}lr@{.}lr@{.}lr@{.}l}
\hline
\hline
 & Comparison & \multicolumn{2}{p{1.5cm}}{Estimate (log odds)}& \multicolumn{2}{c}{Odds}& \multicolumn{2}{p{1.5cm}}{Standard Error} &\multicolumn{2}{c}{$t$-value}& \multicolumn{2}{c}{\emph{p} (Estimated)}\\
\hline
\multirow{2}{3cm}{Base: Control} & SQILabs 1 & 1&10 & 3&00 & 0&28 & 3&18 & &001$^{**}$\\
&SQILabs 2 & 1&51 & 4&50 & 0&33 & 4&58 & $<$&001$^{***}$\\
\hline
\multirow{2}{3cm}{Base: SQILabs 1} & Control & -1&10 & 0&33 & 0&35& -3&18 & &001$^{**}$\\
&SQILabs 2 & 0&41 & 1&50 & 0&29 & 1&39 & &166\\
\hline
\hline
\end{tabular}
\caption{Ordinal logistic regression table for students' follow-up on comparisons. Base comparison is the control group (in the top section) and SQILabs 1 (in the bottom section). $p$-values are estimated from the normal distribution, assuming infinite degrees of freedom, and should be considered a potentially biased estimate. $.$ \emph{p}$<$.1. $^{*}$ \emph{p}$<$.05. $^{**}$ \emph{p}$<$.01. $^{***}$ \emph{p}$<$.001.}
\label{tab:FURegress}
\end{table*}

\vspace{-1em}\subsection{Whether students made changes to their measurements}\label{sec:ChangeIssues}

\vspace{-1em}As outlined in the primary manuscript, convergence issues cause us to split the analyses into two sections. We first describe those issues and how they led to this solution and then present the results from the two analyses.

\vspace{-1em}\subsubsection{Issues with regression for Changes analysis}
\vspace{-1em}We encountered a few problems when performing regression analysis for students' iteration behaviors. The data in Fig. \ref{fig:Changes} suggest a three-way interaction between measurement, group, and whether comparisons were made. The repeated-measures logistic regression model with this three way interaction failed to reach sufficient convergence, presumably due to the low numbers in the top right quadrant of Fig. \ref{fig:Changes} (changes made to Total Internal Reflection by students who did not compare). Stepwise regression models were tested using all appropriate variable selection:

\vspace{-1em}\begin{itemize}
\item Three-way interaction model:  $Measurement \times Group \times Comparisons$ (AIC = 524.7, fails to reach sufficient convergence)
\item Two-way interaction model (1): 
$Measurement \times Group + Group \times Comparisons$ 
(AIC = 524.9, fails to reach sufficient convergence)
\item Two-way interaction model (2): 
$Measurement \times Group + Measurement \times Comparisons$ 
(AIC = 536)
\item Single interaction with Comparisons model: 
$Measurement \times Group + Comparisons$ 
(AIC = 534)
\item Single interaction without Comparisons model: 
$Measurement \times Group$ 
(AIC = 532)
\end{itemize}
\vspace{-0.5em}
\noindent The three-way interaction model produced the best quality fit, as measured by the smallest AIC value. We provide the results of the three-way interaction model in Table \ref{tab:ChangeRegress}. The convergence issues place limitations on our interpretation of the results, however. We, therefore, opted to perform the analysis for whether students did or did not make comparisons separately, assuming the lack of convergence was due to the low sample sizes in the top right quadrant of Fig. \ref{fig:Changes}. We provide the regression analyses for completeness. 

\vspace{-1em}\begin{table*}
\centering
\begin{tabular}{cp{6cm}r@{.}lr@{.}lr@{.}lr@{.}lr@{.}l}
\hline
\hline
 & Comparison & \multicolumn{2}{p{1.5cm}}{Estimate (log odds)}& \multicolumn{2}{c}{Odds}& \multicolumn{2}{p{1.5cm}}{Standard Error} &\multicolumn{2}{c}{$z$-value}& \multicolumn{2}{c}{\emph{p}}\\
\hline
\multirow{6}{3cm}{Logistic Regression (Base: Control [Group]; Snell's Law [Measurement]; No Comparison [Comparisons]} & (Intercept) & 2&16 & 8&67 & 0&58 & 3&73 &$<$ &001$^{***}$\\
&Total Internal Reflection & -5&67 & 0&003 & 1&00 & -5&66 & $<$&001$^{***}$\\
&SQILabs 1 & -0&57 & 0&57 & 0&75 & -0&75 & &447\\
&SQILabs 2 & 13&85 & 1035091&00 & 2987&58 & 0&005 & &996\\
&Explicit Comparison & -1&61 & 0&20 & 0&69 & -2&32 & &020$^{*}$\\
&Total Internal Reflection * SQILabs 1 & 2&22 & 8&69 & 1&17 & 1&89 & &058 . \\
&Total Internal Reflection * SQILabs 2 & 4&26 & 70&81 & 3106&80 & 0&001 & &999\\
&Total Internal Reflection * Explicit Comparison & 2&23 & 9&30 & 1&26 & 1&78 & &075 .\\
&Explicit Comparison * SQILabs 1 & 3&69 & 40&04 & 1&15 & 3&20 & &001$^{**}$\\
&Total Internal Reflection * Explicit Comparison * SQILabs 1 & -3&80 & 0&02 & 1&63 & -2&341 & &019$^{*}$\\
&Total Internal Reflection * Explicit Comparison * SQILabs 2 & -3&58 & 0&03 & 3106&80 & -0&001 & &999\\

\hline\hline
 & Variable & \multicolumn{2}{c}{$\chi^2$} & \multicolumn{2}{p{1.5cm}}{Degrees of freedom} &\multicolumn{2}{c}{$p$}&&\\
\hline
& Measurement & 146&66 & \multicolumn{2}{c}{2} & $<$&001$^{***}$\\
ANOVA & Group & 13&74 & \multicolumn{2}{c}{3} & &003$^{**}$\\
 & Comparisons & 0&10 & \multicolumn{2}{c}{1} & &758\\
& Measurement * Group & 3&54 & \multicolumn{2}{c}{2} & &170\\
& Measurement * Comparison & 0&001 & \multicolumn{2}{c}{1} & &966\\
& Comparison * Group & 4&90 & \multicolumn{2}{c}{2} & &086 . \\
& Measurement * Comparison * Group & 5&47 & \multicolumn{2}{c}{2} & &065 .\\
\hline
\hline
\end{tabular}
\caption{Logistic regression table for whether students made changes to their Snell's Law and Total Internal Reflection measurements. Base comparison is the control group (for year) and Snell's Law (for measurement). This regression failed to reach appropriate convergence and so has limitations on interpretability. $.$ \emph{p}$<$.1. $^{*}$ \emph{p}$<$.05. $^{**}$ \emph{p}$<$.01. $^{***}$ \emph{p}$<$.001.}
\label{tab:ChangeRegress}
\end{table*}

\vspace{-1em}\subsubsection{Changes analysis for students who did not make comparisons}

\vspace{-1em}Table \ref{tab:NoCompChangeRegress} shows the results of the logistic regression analysis that evaluated whether students made changes to the Snell's Law measurement as a function of group, using only students with initially inaccurate measurements and who did not make comparisons. Only the Snell's Law measurement was evaluated because too few students made changes to the Total Internal Reflection measurements.

\vspace{-1em}\begin{table*}
\centering
\begin{tabular}{ccr@{.}lr@{.}lr@{.}lr@{.}lr@{.}l}
\hline
\hline
 & Comparison & \multicolumn{2}{p{1.5cm}}{Estimate (log odds)}& \multicolumn{2}{c}{Odds}& \multicolumn{2}{p{1.5cm}}{Standard Error} &\multicolumn{2}{c}{$z$-value}& \multicolumn{2}{c}{\emph{p}}\\
\hline
\multirow{3}{3cm}{Logistic Regression (Base: Control [Group]} & (Intercept) & 1&42 & 4&14 & 0&42 & 3&38 &$<$ &001$^{***}$\\
&SQILabs 1 & -0&23 & 0&80 & 0&55 & -0&41 & &682\\
&SQILabs 2 & 14&14 & 1383324&00 & 1455&40 & 0&01 & &992\\
\hline\hline
 & Variable & \multicolumn{2}{c}{$\chi^2$} & \multicolumn{2}{p{1.5cm}}{Degrees of freedom} &\multicolumn{2}{c}{$p$}&\multicolumn{4}{c}{}\\
\hline
ANOVA & Group& 0&65 & \multicolumn{2}{c}{2} & &722 & \multicolumn{4}{c}{}\\
\hline
\hline
\end{tabular}
\caption{Logistic regression and ANOVA table for whether students made changes to their Snell's Law measurements, examining only students with initially inaccurate measurements and those that did not make comparisons. Base comparison is the control group (for year). $.$ \emph{p}$<$.1. $^{*}$ \emph{p}$<$.05. $^{**}$ \emph{p}$<$.01. $^{***}$ \emph{p}$<$.001.}
\label{tab:NoCompChangeRegress}
\end{table*}

\vspace{-1em}\subsubsection{Changes analysis for students who did make comparisons}

\vspace{-1em}Table \ref{tab:CompChangeRegress} shows the results of the repeated-measures logistic regression analysis that evaluated whether students made changes to the Snell's Law and Total Internal Reflection measurements as a function of group, using only students with initially inaccurate measurements and who did make comparisons. 

The choice of base reference group in the regression affected the significance of an interaction. That is, the interaction model: $Group \times Measurement$, with the Control group as the base reference produced a non-significant interaction, but the same regression with the SQILabs 1 group as the base reference produced a significant interaction. The simpler model was chosen to decrease the chances of over-fitting. The final model was chosen based on the BIC value. The interaction model produced a lower AIC but higher BIC than the additive model: $Group + Measurement$. Because BIC is more critical of over-fitting, we opted for the simpler model with better BIC.

\vspace{-1em}\begin{table*}
\centering
\begin{tabular}{ccr@{.}lr@{.}lr@{.}lr@{.}lr@{.}l}
\hline
\hline
 & Comparison & \multicolumn{2}{p{1.5cm}}{Estimate (log odds)}& \multicolumn{2}{c}{Odds}& \multicolumn{2}{p{1.5cm}}{Standard Error} &\multicolumn{2}{c}{$z$-value}& \multicolumn{2}{c}{\emph{p}}\\
\hline
\multirow{4}{4cm}{Logistic Regression (Base: Control [Group] \& Snell's Law [Measurement]} & (Intercept) & 0&54 & 1&72 & 0&38 & 1&43 &&152\\
&Total Internal Reflection & -3&42 & 0&03 & 0&60 & -5&67 & $<$&001$^{***}$\\
&SQILabs 1 & 2&00 & 7&40 & 0&51 & 3&94 & $<$&001$^{***}$\\
&SQILabs 2 & 1&80 & 6&06 & 0&50 & 3&60 & $<$&001$^{***}$\\
\hline
\multirow{4}{4cm}{Logistic Regression (Base: Control [SQILabs 1] \& Snell's Law [Measurement]} & (Intercept) & 2&55 & 12&73 & 0&50 & 5&13 &$<$&001$^{***}$\\
&Total Internal Reflection & -3&42 & 0&03 & 0&60 & -5&67 & $<$&001$^{***}$\\
&Control & -2&00 & 0&14 & 0&51 & -3&94 & $<$&001$^{***}$\\
&SQILabs 2 & -0&20 & 0&82 & 0&40 & -&51 & &615\\
\hline\hline
 & Variable & \multicolumn{2}{c}{$\chi^2$} & \multicolumn{2}{p{1.5cm}}{Degrees of freedom} &\multicolumn{2}{c}{$p$}&&\\
\hline
\multirow{2}{*}{ANOVA} & Group & 17&00 & \multicolumn{2}{c}{2} & $<$&001$^{***}$\\
& Measurement & 32&13 & \multicolumn{2}{c}{1} & $<$&001$^{***}$\\
\hline
\hline
\end{tabular}
\caption{Logistic regression and ANOVA table for whether students made changes to their measurements, examining only students with initially inaccurate measurements and those who did make comparisons. Base comparison is the control group (for year) and Snell's Law (for measurement).  $.$ \emph{p}$<$.1. $^{*}$ \emph{p}$<$.05. $^{**}$ \emph{p}$<$.01. $^{***}$ \emph{p}$<$.001.}
\label{tab:CompChangeRegress}
\end{table*}

\end{appendix}
\end{document}